\documentclass[aps,prl,showpacs,twocolumn]{revtex4}
\usepackage{epsfig,graphicx}

\input{tcilatex}

\begin{document}

\title{Pairwise Entanglement and Local Polarization of Heisenberg Model}
\author{Xiao-Qiang Xi,$^{1,2}$ R. H. Yue$^3$ and W. M. Liu$^2$}
\address{$^1$Department of Applied Mathematics and Physics,
 Xi'an Institute of Posts and Telecommunications, Xi'an 710061, China}
\address{$^2$Beijing National Laboratory for Condensed Matter Physics,
Institute of Physics, Chinese Academy of Sciences, Beijing 100080,
China}
\address{$^3$Institute of Modern Physics, North-West
University, Xi'an 710069, China}

\begin{abstract}
The pairwise entanglement and local polarization of the ground state
are discussed by studying the Heisenberg XX model in
 finite qubit case. The results show that: the ground state is
 composed by the micro state with the minimal total spin 0
 (for even qubit) or $\frac{1}{2}$ (for odd qubit), local
 polarization (LP) has intimate relation with the probability of the
 micro state in the ground state, the stronger the LP the smaller the probability, the same
LP corresponding to the same probability; the pairwise entanglement
of the ground state is the biggest in all the eigenvectors. We find
when the qubit is small, the degenerate of state will decrease the
pairwise entanglement, there has great different between the odd and
the even qubit chain; when the qubit number is big, the effect of
qubit number to the pairwise entanglement will disappear, the
limited value will be round about 0.3424.
\end{abstract}


\maketitle

Quantum entanglement play an important role in modern physics,
besides using to test some fundamental questions of the quantum
mechanics and using in quantum information processing, it can be
used in the sensitivity of interferometric measurements such as
quantum lithography \cite{BotoPRL85_2733}, quantum optical gyroscope
\cite{DowlingPRA57_4736}, quantum clock synchronization and
positioning \cite{JozsaPRL85_2010} and frequency metrology
\cite{BollingerPRA54R4649}, it also play a central role in the study
of strongly correlated quantum systems \cite{PreskillJMO47_127},
especially, ground-state entanglement has some connection with the
quantum phase transition \cite{OsbornePRA66_032110}, Mott
insulator-superfluid transition and quantum magnet-paramagnet
transition. Those usability are all based on the production of
entanglement, which is dependent on our recognition to it, because
we know the two level system and their interaction are the necessary
condition for the pairwise entanglement, so photons
 \cite{Bouwmeester97N390}, trapped ions \cite{Turchette98PRL81} (or
trapped atoms \cite{Rauschenbeutel00S288}, crystal lattices
 \cite{Yamaguchi99APA68}, Josephson junctions \cite{Makhlin99N398} and
Bose-Einstein condensates \cite{Sorensen01N409}, which have two
level system and interaction, are often used to produce
entanglement. Looking for the law of entanglement become one of the
most important thing.

Spin chain is a nature candidate for producing pairwise
entanglement, it has been used to construct a quantum computer
\cite{Loss9899}, C-NOT gate \cite{Imamoglu1999} and swap gate
\cite{Wang2001PR}; it can also be used to quantum communication
\cite{PRL91_207901,PRA69_034304,PRA69_052315}. For the pairwise
entanglement in spin chain, the relevant work are around how to
control and mike it maximal, the effective factors are temperature,
interchange coupling \cite{Arnesen2000,Nielsen2000,Wang2001PR},
magnetic field and system impurity
\cite{Wang0105,XiPLA297,AhmadPRA052105}, some useful conclusions are
concluded for the few qubits cases. Besides those factors, the
length of chain is also an effective factor to the pairwise
entanglement, the previous work in this point are some special
cases: for $|W_{N}>$ state, the concurrences between any two qubits
are all equal to $2/N$ \cite{PRA62_062314,PRL80_2245}; Koashi {\em
et al.} show \cite{PRA62_050302} that the maximum degree of
entanglement between any pair of qubits of a N-qubit symmetric state
is $2/N$; Connor {\em et al.} \cite{PRA63_052302} calculated the
pairwise entanglement in spin chain which satisfied two conditions:
the state $|\psi>$ of the ring is an eigenvector of the total
z-component of spin; neighboring particles cannot both be in the
state $|\uparrow>$. Those work are contribute to our understanding
of the properties of entanglement and its distributed among many
objects, but even the work about pairwise entanglement is far from
completeness, for example, how the macro quality, such as total spin
, LP and the length of chain, effect the pairwise entanglement?

In this paper, we will use the simplest spin model (Heisenberg XX
model) to study how the total spin, LP and the length of chain
effect the pairwise entanglement. The results can help us to deep
understand the properties of entanglement and give some useful
guidance to the experiments in solid system.

No one knows if nature only needs pairwise entanglement, but it is
no doubt that it is very useful in quantum information processing,
and it is necessary to find all its properties. Before beginning,
let us give a brief review of the measurement of the pairwise
entanglement and the notation in this paper: the concept of
entanglement of formation and concurrence
\cite{Wootters1998,Hill1997}. Concurrence ${C}$ range from zero to
one and it is monotonically relate to entanglement of formation, so
that concurrence $C$ is a kind of measure of entanglement. $\rho$ is
the density matrix of system, ${\rho}_{12}= Tr_{non(12)} \rho$
(mixed or pure) is the reduced density matrix of the pair (the
nearest or the non-nearest), where $1,2$ mean the first qubit and
the second qubit in the Heisenberg XX ring. The concurrence
corresponding to the density matrix is defined as ${C}_{12}=
\mbox{max} \{ {\lambda}_{1} -{\lambda}_{2} -{\lambda}_{3}
-{\lambda}_{4},0 \}$, where $\lambda_k$, $k=1,2,3,4$ are the square
roots of the eigenvalues of the operator ${\hat{\rho}}_{12}=
{\rho}_{12} ({\sigma}_1^y{\otimes}{\sigma}_2^y) {\rho}_{12}^*
({\sigma}_1^y{\otimes}{\sigma}_2^y)$ in descending order.

The Hamiltonian of $S=\frac{1}{2}$ Heisenberg XX chain is
$H_{xx}=\sum_{n=1}^{N}{2J(S_n^xS_{n+1}^x+S_n^yS_{n+1}^y)}$, where
$S_{N+1}=S_{1}$, $J<0$ and $J>0$ are corresponding to ferromagnetic
and anti-ferromagnetic area. Using the relations
$S^{\alpha}={\sigma}^{\alpha}/2\ (\alpha=x,y)$ and
${\sigma}^{\pm}=({\sigma}^x \pm i{\sigma}^y)/2$, the system
Hamiltonian can be rewritten as
$H=J\sum_{n=1}^{N}({\sigma}_n^+{\sigma}_{n+1}^-+{\sigma}_{n+1}^+{\sigma}_{n}^-)$.

{\em The eigenvalues and eigenvectors.-} The most important step of
calculating concurrence $C_{12}$ is to get the eigenvalues and
eigenvectors of the system. Usually, suppose
$|\psi>=\sum_{i=0}^{2^N-1}a_i|i>$ and using
 $H|\psi>=E|\psi>$ we can get
all eigenvalues and eigenvectors of the system, then pick out the
ground state in them. For $N$-qubit case, we must handle a matrix
with order $2^N$, the calculation will be very difficult when $N$ is
large.

In order to overcome the difficulty of calculation, we find a way to
reduce the coefficients in the eigenvectors. As we know, in four and
five qubit case, the ground sates of the ferromagnetic and
anti-ferromagnetic area are all composed by the micro state which
has minimal total spin $\frac{1}{2}$ (for odd qubit) or 0 (for even
qubit); the micro states, which have the same macro state, have the
same probability. So we generalize this to more qubit case, the
results show that this method is still effective. Details are shown
in the following:

For three qubit chain, the ground state is composed by $|j>$ , or
$|j,j+1>$, where $j=1,2,3$ is a space coordinate and at this site
the spin is up, the others are spin down. All the three $|j>$ has
the same probability, so we can write the eigenvector as
$|\psi(m)>=C_{1}\sum_{j=1}^{3}e^{i\frac{2j\pi}{3}m}|j>$, where
$m=1,2,3$, $C_{1}$ satisfy $H|\psi>=E|\psi>$ and $3|C_{1}|^2=1$. In
the ferromagnetic area $(J<0)$, the eigenvalue is $E_3^{+}=2J$, the
corresponding eigenvectors(two degenerate) are
$|\psi(3)>_{1}^{+}=\frac{1}{\sqrt{3}}\sum_{j=1}^{3}|j>, \
|\psi(3)>_{2}^{+}=\frac{1}{\sqrt{3}}\sum_{j=1}^{3}|j,j+1>.$ In the
anti-ferromagnetic area $(J>0)$, the eigenvalue is $E_3^{-}=-J$, the
corresponding eigenvectors(four degenerate) are $|\psi(1)>_{1}^{-}
=\frac{1}{\sqrt{3}}\sum_{j=1}^{3}e^{i\frac{2j\pi}{3}}|j>, \
|\psi(2)>_{2}^{-}=\frac{1}{\sqrt{3}}\sum_{j=1}^{3}e^{i\frac{2j\pi}{3}2}|j>,
\
|\psi(1)>_{3}^{-}=\frac{1}{\sqrt{3}}\sum_{j=1}^{3}e^{i\frac{2j\pi}{3}}|j,j+1>,
\
|\psi(2)>_{4}^{-}=\frac{1}{\sqrt{3}}\sum_{j=1}^{3}e^{i\frac{2j\pi}{3}2}|j,j+1>.$

For four qubit chain, the ground state is composed by $|j,j+1>$
(i.e.the nearest two neighbours spin up) and $|j,j+2>$ (the next
nearest two neighbours spin up), where $j=1,2,3,4$.
 We write the eigenvector of four qubit case as
$|\psi(m_i)>=C_{1}\sum_{j=1}^{4}e^{i\frac{2j\pi}{4}m_1}|j,j+1>
+C_{2}\sum_{j=1}^{2}e^{i\frac{2j\pi}{2}m_2}|j,j+2>$, where
$m_1=1,2,3,4,m_2=1,2$, $C_{1},C_{2}$ satisfy $H|\psi>=E|\psi>$ and
$4|C_{1}|^2+2|C_{2}|^2=1$. The eigenvalues in both magnetic area are
$E_4^{+}=2\sqrt{2}J\ (J<0),\ E_4^{-}=-2\sqrt{2}J\ (J>0)$, the
corresponding eigenvector has no degenerate in both magnetic areas,
they are $|\psi(4,2)>^{+}=
\frac{1}{2\sqrt{2}}\sum_{j=1}^{4}|j,j+1>+\frac{1}{2}\sum_{j=1}^2|j,j+2>,\
|\psi(4,2)>^{-}=
\frac{1}{2\sqrt{2}}\sum_{j=1}^{4}|j,j+1>-\frac{1}{2}\sum_{j=1}^2|j,j+2>$.

For five qubit chain, the ground state is composed by $|j,j+1>$ and
$|j,j+2>$, or $|j,j+1,j+2>$ and $|j,j+2,j+3>$, where $j=1,2,3,4,5$.
The eigenvector is
$|\psi(m_i)>=C_{1}\sum_{j=1}^{5}e^{i\frac{2j\pi}{5}m_1}|j,j+1>
+C_{2}\sum_{j=1}^{5}e^{i\frac{2j\pi}{5}m_2}|j,j+2>$, where
$m_1,m_2(=1,2,3,4,5)$, $C_{1},C_{2}$ satisfy $H|\psi>=E|\psi>$ and
$5|C_{1}|^2+5|C_{2}|^2=1$. In the ferromagnetic area, the eigenvalue
is $E_5^{+}=(\sqrt{5}+1)J$, the corresponding eigenvectors(two
degenerate) are $|\psi(5,5)>_{1}^{+}=0.235|j,j+1>+0.380|j,j+2>,\
|\psi(5,5)>_{2}^{+}=0.235 |j,j+1,j+2>+0.380|j,j+2,j+3>$, the
eigenvalue is $E_5^{-}=-\frac{1}{2}(3+\sqrt{5})J$, the corresponding
eigenvectors(four degenerate) are
$|\psi(1,1)>_{1}^{-}=C_{11}\sum_{j=1}^{5}e^{i\frac{2j\pi}{5}}
|j,j+1>+C_{12}\sum_{j=1}^{5}e^{i\frac{2j\pi}{5}}|j,j+2>,\
|\psi(4,4)>_{2}^{-}=C_{21}\sum_{j=1}^{5}e^{i\frac{2j\pi}{5}4}
|j,j+1>+C_{22}\sum_{j=1}^{5}e^{i\frac{2j\pi}{5}4} |j,j+2>,\
|\psi(1,1)>_{3}^{-}=C_{31}\sum_{j=1}^{5}e^{i\frac{2j\pi}{5}}
|j,j+1,j+2>+C_{32}\sum_{j=1}^{5}e^{i\frac{2j\pi}{5}} |j,j+2,j+3>,\
|\psi(4,4)>_{4}^{-}=C_{41}\sum_{j=1}^{5}e^{i\frac{2j\pi}{5}4}
|j,j+1,j+2>+C_{42}\sum_{j=1}^{5}e^{i\frac{2j\pi}{5}4}|j,j+2,j+3>.$
where parameters $C_{ij}$ are: $C_{11}=-0.190+0.138i,\
C_{21}=-0.190-0.138i,\ C_{31}=0.073+0.224i,\ C_{41}=0.073-0.224i,\
C_{12}=C_{22}=C_{32}=C_{42}=0.380$.

For six qubit chain, the ground state is composed by $|j,j+1,j+2>$,
$|j,j+2,j+3>$, $|j,j+3,j+4>$ and $|j,j+2,j+4>$, where
$j=1,2,3,4,5,6$, the eigenvector is
$|\psi(m_i)>=C_{1}\sum_{j=1}^{6}e^{i\frac{2j\pi}{6}m_1}|j,j+1,j+2>
+C_{2}\sum_{j=1}^{6}e^{i\frac{2j\pi}{6}m_2}|j,j+2,j+3>$
$+C_{3}\sum_{j=1}^{6}e^{i\frac{2j\pi}{6}m_3}|j,j+3,j+4>,
+C_{4}\sum_{j=1}^{2}e^{i\frac{2j\pi}{2}m_4}|j,j+2,j+4>$ where
$m_{1,2,3}=1,2,3,4,5,6$,$m_4=1,2$, $C_{1},C_{2},C_{3},C_{4}$ are
determined from $H|\psi>=E|\psi>$ and
$6|C_{1}|^2+6|C_{2}|^2+6|C_{3}|^2+2|C_{4}|^2=1$. The eigenvalues in
both magnetic area are $E_6^{+}=4J\ (J<0),\ E_6^{-}=-4J\ (J>0)$, the
corresponding eigenvector has no degenerate in both magnetic areas,
they are $|\psi(6,6,6,2)>^{+}=\frac{1}{6\sqrt{2}}\sum_{j=1}^{6}
|j,j+1,j+2>+\frac{1}{3\sqrt{2}}\sum_{j=1}^{6} |j,j+2,j+3>
+\frac{1}{3\sqrt{2}}\sum_{j=1}^{6}
|j,j+3,j+4>+\frac{1}{2\sqrt{2}}\sum_{j=1}^{2} |j,j+2,j+4>;\
|\psi(3,3,3,1)>^{-}=\frac{1}{6\sqrt{2}}\sum_{j=1}^{6}e^{i\frac{2j\pi}{6}3}
|j,j+1,j+2>+\frac{1}{3\sqrt{2}}\sum_{j=1}^{6}e^{i\frac{2j\pi}{6}3}
|j,j+2,j+3>+\frac{1}{3\sqrt{2}}\sum_{j=1}^{6}e^{i\frac{2j\pi}{6}3}
|j,j+3,j+4>+\frac{1}{2\sqrt{2}}\sum_{j=1}^{2}e^{i\frac{2j\pi}{2}}|j,j+2,j+4>$.

For more than six qubit chain, we calculated the case of seven and
eight qubit, the eigenvalue and eigenvectors of seven qubit in
ferromagnetic (anti-ferromagnetic is omitted) area are:
$E_7=4.4611J,\ |\psi>_1=\sum_{kl}(C_{kl}\sum_{j=1}^{7}|j,j+k,j+l>),\
|\psi>_2=\sum_{kl}(C_{kl}\sum_{j=1}^{7}|\bar{j},\ \overline{j+k},\
\overline{j+l}>)$, where $kl=12,13,14,15,24$ is composed number,
$C_{12,13,14,15,24}=0.064,0.143,0.178,0.143,0.257$, $\bar{j}$ means
the spin in this site is down. The eigenvalue and eigenvectors of
eight qubit are: $E_8=2\sqrt{4+2\sqrt{2}}J,|\psi>
=\sum_{kln}(C_{kln}\sum_{j=1}^{8}|j,j+k,j+l,j+n>)
+C_{145}\sum_{j=1}^4|j,j+1,j+4,j+5>+C_{246}\sum_{j=1}^2|j,j+2,j+4,j+6>$,
where $kln=123,124,125,126,134,135,136,146$ is also composed number,
$C_{123,124,125,126,134,135,136,146}=0.022,0.056,0.074,0.056,0.074,
0.136,0.127,0.136,C_{145}=0.147,C_{246}=0.417$. Using this ways the
ground state of $N$ qubit can be calculated as long as we can
manufacture a matrix of order
$\frac{(N-1)!}{[\frac{N}{2}]!(N-[\frac{N}{2}])!}$, in fact only the
case of $N\le15$ (the matrix order of $N=15$ is 429, great less than
$2^{15}=32768$) can be calculate. However, $N=15$ is large for the
quantum dot or the Josephson junctions, which can realize the spin
system.

The degenerate degree of the ground state is different for odd and
even qubit chain: odd qubit chain has degenerate while even qubit
case has not. For odd qubit case, different magnetic area has
different degenerate, the degenerate in anti-ferromagnetic area is
higher than ferromagnetic area. We can understand them from the
following: (1) for the minimal total spin of even (N) qubit chain,
they only have one possibility of $\frac{N}{2}$ spin up, of cause
they have no degenerate; (2) for the minimal total spin of odd(N)
qubit chain, they have two equal possibility, $\frac{N-1}{2}$ or
$\frac{N+1}{2}$ spin up, so it at least has two degenerate, in
 the anti-ferromagnetic area, when two spin exchange, a minus will
appear, so it has four degenerate. This degenerate will decrease the
pairwise entanglement. The degenerate will decrease the pairwise
entanglement.

{\em Local Polarization (LP).-} From the above eigenvectors, we can
see a very important phenomena about LP. We summarized the micro
state of even qubit into Table 1.

{\footnotesize{Table 1. The LP and probability of micro state,
"P-Mic" is instead of the probability of the micro state}
\begin{center}
{\footnotesize
\begin{tabular}{|c|c|c|}
\hline $N$ & Micro state &P-Mic \cr \hline
4-qubit&$|j,j+1>$&$\frac{1}{8}$ \cr
  & $|j,j+2>$&$\frac{1}{4}$ \cr \hline
6-qubit &$|j,j+1,j+2>$ &$\frac{1}{72}$ \cr
        &$|j,j+2,j+3>$ &$\frac{1}{18}$ \cr
        &$|j,j+3,j+4>$ &$\frac{1}{18}$ \cr
        &$|j,j+2,j+4>$ &$\frac{1}{8}$ \cr \hline
8-qbit  &$|j,j+1,j+2,j+3>$ &$0.022^2$ \cr
        &$|j,j+1,j+2,j+4>$ &$0.056^2$ \cr
        &$|j,j+1,j+2,j+6>$ &$0.056^2$ \cr
        &$|j,j+1,j+2,j+5>$ &$0.074^2$ \cr
        &$|j,j+1,j+3,j+4>$ &$0.074^2$ \cr
        &$|j,j+1,j+3,j+6>$ &$0.127^2$ \cr
        &$|j,j+1,j+3,j+5>$ &$0.136^2$ \cr
        &$|j,j+1,j+4,j+6>$ &$0.136^2$ \cr
        &$|j,j+1,j+4,j+5>$ &$0.147^2$ \cr
        &$|j,j+2,j+4,j+6>$ &$0.417^2$ \cr \hline
\end{tabular}}
\end{center}}

Now let us give an analysis of Table 1 in details. In four qubit,
the components are $|j,j+1>$ and $|j,j+2>$, the former has big LP
and small probability, while on the country for the later. The
components of six qubit are $|j,j+1,j+2>,|j,j+1,j+3>,|j,j+1,j+4>$
and $|j,j+2,j+4>$, the corresponding probability of them are
$\frac{1}{72},\frac{1}{18},\frac{1}{18},\frac{1}{8}$, $|j,j+1,j+2>$
is the smallest, $|j,j+2,j+4>$ is the biggest, the middle two are
equal. Obviously, the LP of them are very different, $|j,j+1,j+2>$'s
is the strongest (with high degree of order), $|j,j+2,j+4>$'s is the
weakest (with low degree of order), the LP of the middle two are
equal (they have the same probability). In eight qubit, for
simplicity we use the composed number $kln$ instead of the micro
state (for example, 123 instead of $|j,j+1,j+2,j+3>$), 123 has the
strongest LP and the smallest probability, 246 has the weakest LP
and the biggest probability; there have three same pairs in the
middle, they are 124 and 126, 125 and 134, 135 and 146, they have
the same LP and probability. The total spin of them (even qubit) are
all equal zero. Similarly phenomena exists in odd qubit cases (total
spin equal $\frac{1}{2}$).

That is to say, the LP and the probability of micro state have point
to point relation: the stronger the LP the smaller the probability,
the same LP corresponding to the same probability. The micro states
with the biggest LP has no pairwise entanglement, the micro states
with the smallest LP has the maximal pairwise entanglement between
any two qubits. The other's are in the middle of two extremum. They
compose the stablest state, no more or no less, in some degree, spin
chain shows harmonic and variety, just like nature as.

{\em The Pairwise Entanglement.-} Using the standard concurrence
theory, we can get the pairwise entanglement of the ground state as
the following, see Table 2:

{\footnotesize {Table 2. The table of $C_{12}$ at ground state, "f"
and "a-f" are instead of ferromagnetic and anti-ferromagnetic area.}
\begin{center}
{\footnotesize
\begin{tabular}{|c|c|c|c|c|c|c|}
\hline $N$ & &2-qubit&4-qubit&6-qubit&8-qubit\cr \hline
${{\cal{C}}_{12}}$ & $f$   &1 &0.45711&0.38889&0.37048 \cr \hline
${{\cal{C}}_{12}}$ & $a-f$ &1 &0.45711&0.38889&0.37048\cr
\hline\hline $N$& &1-qubit&3-qubit&5-qubit&7-qubit\cr \hline
${{\cal{C}}_{12}}$ & $f$   & &0.33333&0.33666&0.33787\cr \hline
${{\cal{C}}_{12}}$ & $a-f$ & &0&0.21305& 0\cr \hline
\end{tabular}}
\end{center}}

From Table 2, we find that $C_{12}$ (the concurrence between the
nearest qubits) is the same at different ferromagnetic area for even
qubit chain, while different for odd qubit chain; for odd qubit
chain, $C_{12}$ at anti-ferromagnetic area is bigger than
ferromagnetic area. These different come form the different of
degenerate degree: even qubit chain has no degenerate while odd
qubit chain has; for odd qubit chain, the degenerate degree at
anti-ferromagnetic area is double times than that of ferromagnetic
area. Those can be seen clearly from eigenvectors and can be
testified by calculating $C_{12}$ of any single state. The bigger
the degenerate the smaller the entanglement, so decreasing the
degenerate degree will increasing pairwise entanglement, introducing
magnetic field is a proper ways to eliminate the degenerate and
increase the entanglement, for example, if we eliminate the
degenerate in the ground state of three-qubit case, we will get
$C_{12}=0.6667$ in the ferromagnetic area. The pairwise entanglement
in anti-ferromagnetic area of odd qubit chain seems complex, because
there exist entanglement in 5-qubit chain while do not exist in
3-qubit and 7-qubit chain, it worth to be discussed further.

From Table 2, we see that $C_{12}$ decreases with qubit number for
even qubit chain while on the country for odd qubit chain. For
understanding this more clearly, we plot the diagram of $C_{12}$
with qubit number at ferromagnetic area, see Figure 1:
\begin{center}{\epsfxsize 48mm \epsfysize 36mm \epsffile{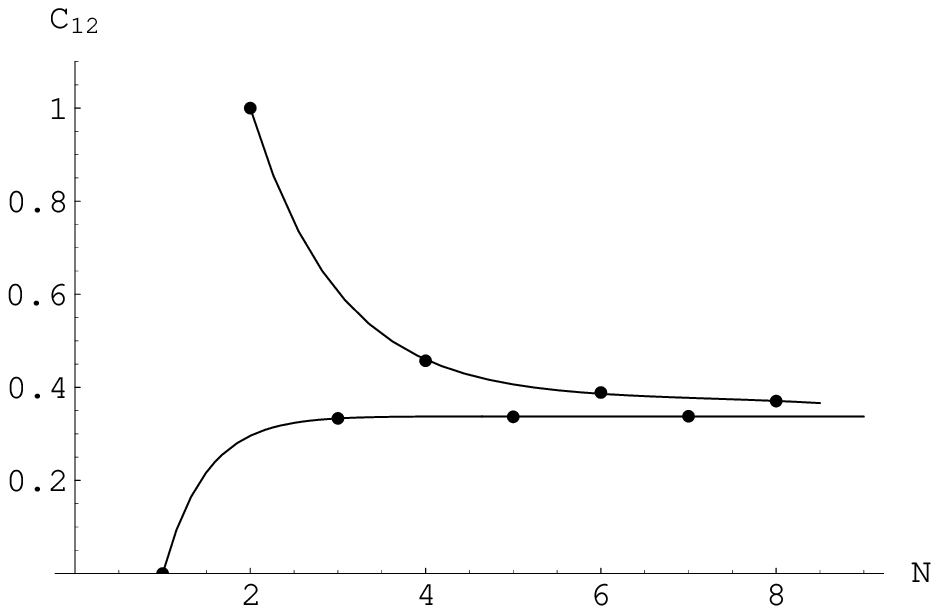}}
\end{center}
{\footnotesize Figure 1. The diagram of $C_{12}$ with qubit number
at ferromagnetic area}

Obviously, from the trend in the diagram of Figure 1, we may infer
that there exist a limit value of $C_{12}$ and it will be round
about 0.3424, the trend in the anti-ferromagnetic area are similarly
with ferromagnetic area as well as its limit value. That is to say,
as the qubit number increase, the effect of qubit number to the
pairwise entanglement will disappear, this is coincidence with the
fact that single qubit will give smaller and smaller effect to the
system as the qubit number increase. This conclusion tells us if we
want to use such pairwise entanglement, a short chain is enough.

Up to now, we constructed the eigenvectors of the ground state of
 Heisenberg XX chain in finite case, discussed the local polarization (LP)
 and probability of the micro states
and calculated the pairwise entanglement. There are three
interesting results can help us to deep understand the properties of
entanglement and give us some guidance in the future experiments.

 (1) The micro states which compose the ground state have
 the minimal total spin and the same macro state, then they have
 the same probability. In finite qubit case, this way can take
 us great advantage for getting the eigenvalues and eigenvectors.
 We suppose that minimal total spin for a macro state
 is a necessary condition for the ground state, this point is coincidence
 with other law of nature: such as law of minimal energy for a stable system,
 law of minimal superficial area for a drop of liquid and law of
 minimal action for the real orbit of a subject {\em etc.}.

(2) The pairwise entanglement of the ground state is the maximal in
all eigenvectors. Because pairwise entanglement is a kind of
correlated quality, the more correlation between the subsystem, the
more stability there exist in the system. So entanglement can tells
us whether a state is the ground state or not, this is very
interesting and maybe this property will became a criterion for the
ground state, if that, entanglement will have a new usage.

(3) The LP and the probability of the micro state have intimate
relation: the stronger the LP the smaller the probability, the same
LP corresponding to the same probability. The state of the smallest
LP is a special state, any qubit pair (nearest or non-nearest) has
the maximal pairwise entanglement (concurrence=1). This chain is an
ideal quantum entanglement channel, which can be used to teleport a
quantum state and realize a quantum network. The important thing is
how to design such a state in real system. If we can control a
single spin with magnetic field then we can realize such an
eigenvector.

\vskip 2.5mm

This work is supported in part by the NSF of China under grant
10547008, 90403034, 90406017, 60525417, by the National Key Basic
Research Special Foundation of China under 2005CB724508, by the
Foundation of Xi'an Institute of Posts and Telecommunications under
grant 105-0414 and by the NSF of Shanxi Province under grant
2004A15. We also acknowledge the support from CAS through the
project: International Team on Superconductivity and Novel
Electronic Materials (ITSNEM).

\end{document}